\documentclass{article}
\usepackage{spconf,amsmath,graphicx}
\usepackage{bm, bbm}
\usepackage{amssymb,amsfonts}
\usepackage{caption}
\usepackage{subcaption}
\usepackage{siunitx}
\usepackage[hyphens]{url}
\usepackage{standalone,tikz}
\usepackage{textcomp}
\usetikzlibrary{shapes,arrows,calc,fit, decorations.pathreplacing}
  
\usepackage{hyperref}
\usepackage{lipsum}
\newcommand\copyrighttext{%
  \footnotesize \textcopyright 2021 IEEE. Personal use of this material is permitted. Permission from IEEE must be obtained for all other uses, in any current or future media, including reprinting/republishing this material for advertising or promotional purposes, creating new collective works, for resale or redistribution to servers or lists, or reuse of any copyrighted component of this work in other works}
\renewcommand\copyrightnotice{%
\begin{tikzpicture}[remember picture,overlay]
\node[anchor=south,yshift=10pt] at (current page.south) {\fbox{\parbox{\dimexpr\textwidth-\fboxsep-\fboxrule\relax}{\copyrighttext}}};
\end{tikzpicture}%
}
\title{Joint Far- and Near-End Speech Intelligibility Enhancement based on the Approximated Speech Intelligibility Index}
\name{\begin{tabular}{c}Andreas Jonas Fuglsig$^{\star \dagger}$, Jan Østergaard$^{\dagger}$, Jesper Jensen$^{\dagger}$, Lars Søndergaard Bertelsen$^{\star}$,\\ Peter Mariager$^{\star}$, Zheng-Hua Tan$^{\dagger}$\thanks{This work is partly supported by Innovation Fund Denmark Case no. 9065-00204B.}\end{tabular}}

\address{$^{\star}$ RTX A/S, Nørresundby, Denmark\\
$^{\dagger}$Aalborg University, Aalborg, Denmark}
\begin{document}
\ninept
\maketitle
\copyrightnotice
\global\csname @topnum\endcsname 0
\global\csname @botnum\endcsname 0
\begin{abstract}
This paper considers speech enhancement of signals picked up in one noisy environment which must be presented to a listener in another noisy environment. Recently, it has been shown that an optimal solution to this problem requires the consideration of the noise sources in both environments jointly.  However, the existing optimal mutual information based method requires a complicated system model that includes natural speech variations, and relies on approximations and assumptions of the underlying signal distributions. In this paper, we propose to use a simpler signal model and optimize speech intelligibility based on the Approximated Speech Intelligibility Index (ASII). We derive a closed-form solution to the joint far- and near-end speech enhancement problem that is independent of the marginal distribution of signal coefficients, and that achieves similar performance to existing work. In addition, we do not need to model or optimize for natural speech variations. 
\end{abstract}
\begin{keywords}
Multi-microphone, speech intelligibility enhancement, ASII, beamformer
\end{keywords}
\section{Introduction}
\label{sec:intro}

Speech communication systems, such as mobile telephony, hearing aids and intercom systems, are required to work in numerous environments. As a consequence, the user environment is often noisy which can lead to intelligibility problems.

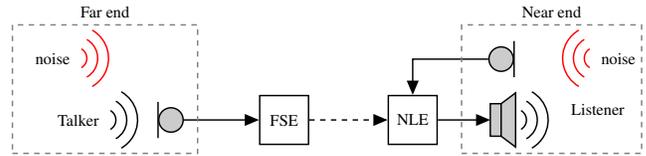
\begin{figure}
    \begin{center}
        % \begin{scaletikzpicturetowidth}{\columnwidth}
        % \end{scaletikzpicturetowidth}
        % \includegraphics[width=\columnwidth]{concept_disjoint_process.pdf}
        % Definition of blocks:
\tikzset{%
  block/.style    = {draw, thick, rectangle, minimum height = 3em,
    minimum width = 3em, align=center},
  sum/.style      = {draw, circle, node distance = 2cm}, % Adder
  input/.style    = {coordinate}, % Input
  output/.style   = {coordinate}, % Output
  cross/.style={path picture={ 
	\draw[black]
  (path picture bounding box.south east) -- (path picture bounding box.north west) (path picture bounding box.south west) -- (path picture bounding box.north east);}},
  circle_gain/.style = {draw, cross, circle, minimum height=2.3em},
  speaker/.pic={
    \filldraw[fill=gray!40,pic actions] 
    (-15pt,0) -- 
      coordinate[midway] (-front) 
    (15pt,0) -- 
    ++([shift={(-6pt,8pt)}]0pt,0pt) coordinate (aux1) -- 
    ++(-18pt,0) coordinate (aux2) 
    -- cycle 
    (aux1) -- ++(0,6pt) -- coordinate[midway] (-back) ++(-18pt,0) -- (aux2);
  },
  radiation/.style={{decorate,decoration={expanding waves,angle=90,segment length=6pt}}}
}
% Defining string as labels of certain blocks.
\newcommand{\suma}{\Large$+$}
\newcommand{\inte}{$\displaystyle \int$}
\newcommand{\derv}{\huge$\frac{d}{dt}$}

\begin{tikzpicture}[auto, thick, node distance=2.5cm, >=triangle 45]
% \draw[help lines] (-3,-4) grid (11,2);
% \draw[step=1mm, gray, thin] (-3,-4) grid (11,2); 
\draw
	% Drawing the blocks of first filter :
	node at (0,0) (source) {Talker}
	node[above of =source, xshift=-.5cm, node distance=1.2cm] (fe_noise1) {noise}
	% node[below of =source, xshift=-.7cm, node distance=.9cm] (fe_noise2) {noise}
	node[right of=source, fill=gray!40, circle, draw, minimum height=1.5em, node distance=1.8cm]  (fe_mic) {}
    ([yshift=10pt] fe_mic.west) -- ([yshift=-10pt] fe_mic.west)
	node [block, right of=fe_mic, node distance=2.2cm] (fle) {FSE}
	node [block, right of=fle] (nle) {NLE}
	;

\pic[right of=nle, rotate=90, node distance=2cm] (ls) {speaker};

\draw
	node[above of=lsaux2, fill=gray!40, circle, draw, minimum height=1.5em, node distance=1.5cm]  (ne_mic) {}
	([yshift=10pt] ne_mic.east) -- ([yshift=-10pt] ne_mic.east)
	node[right of=ls-front, node distance=1.6cm, yshift=.2cm] (listener) {Listener}
	node[right of =ls-front, yshift=1.2cm, node distance=2cm] (ne_noise1) {noise}
	% node[right of =ls-front, yshift=-.8cm, node distance=2cm] (ne_noise2) {noise}
	;
	
    % Joining blocks. 
    % Commands \draw with options like [->] must be written individually
	\draw[thick,radiation,decoration={angle=60}] ([xshift=.0cm]source.east) -- +(0:.7);
	\draw[thick,radiation,decoration={angle=60}] ([xshift=.0cm]ls-front) -- +(0:.7);
	\draw[thick,radiation,decoration={angle=60},color=red] ([xshift=.0cm]fe_noise1.east) -- +(0:.7);
	% \draw[thick,radiation,decoration={angle=60},color=red] ([xshift=.0cm]fe_noise2.east) -- +(0:.7);
	\draw[thick,radiation,decoration={angle=60},color=red] ([xshift=.0cm]ne_noise1.west) -- +(180:.7);
	% \draw[thick,radiation,decoration={angle=60},color=red] ([xshift=.0cm]ne_noise2.west) -- +(180:.7);
	\draw[->](fe_mic) -- (fle);
	\draw[->, dashed](fle) -- (nle);
	\draw[->](nle) -- (ls-back);
	\draw[->](ne_mic) -| (nle);

	\node[color=gray,thick, dashed, draw, rectangle, xshift=.0cm, yshift=.6cm,minimum width=3.6cm, minimum height=2.5cm, label={Far end}] at (source.east) {};
	\node[color=gray,thick, dashed, draw, rectangle, xshift=.7cm, yshift=.6cm,minimum width=3.5cm, minimum height=2.5cm, label={Near end}] at (ls-front) {};

\end{tikzpicture}
        % !!! Blurry in PDF preview only not in Adobe !!!
        \caption{\normalsize Classic speech communication system with Far-end Speech Enhancement (FSE) and Near-end Listening Enhancement (NLE).}\vspace{-6mm}
        \label{fig:concept_disjoint_proc}
    \end{center}
\end{figure}

For speech communication systems we may consider two different environments, cf. Fig.~\ref{fig:concept_disjoint_proc}; the far-end environment (at the target talker) and the near-end environment (at the listener). Both the far- and near-end environment are often noisy, which leads to degradations in both speech quality and Speech Intelligibility (SI) for the listener. 
%At the far-end, interfering acoustic sources get mixed with the target speech, thereby reducing the quality and intelligibility of the signal presented to the listener in the near-end. Furthermore, as the signal is played out in the near-end environment additive noise and/or possible reverberation may lead to further reduced intelligibility and quality for the listener. 
To remedy these effects, speech enhancement techniques may be applied at both the far- and near-end.
% Far-end speech enhancement deals with improvement of quality and intelligibility of speech that has been degraded by additive noise or reverberation. Therefore, 
% Far-end speech enhancement is sometimes referred to as noise reduction~\cite{loizou_speech_2013}. 
Depending on the availability, Far-end Speech Enhancement (FSE) algorithms may utilize either a single or multiple microphones \cite{loizou_speech_2013,gannot_consolidated_2017, hu_comparative_2007,eneman_evaluation_2008}.
In contrast to the far-end scenario, in Near-end Listening Enhancement (NLE)\cite{cooke_listening_2014, kleijn_optimizing_2015}
% the near-end
the interfering noise is mixed with the target speech after processing. Therefore, the noise cannot be reduced by the usual post-processing techniques of the former scenario. Instead, before playback in the noisy environment, NLE increases SI by adaptively pre-processing the FSE signal received from the far-end, while exploiting knowledge about the near-end noise.
%An obvious approach to increasing SI in additive noise environments is to increase the playback level of the output speech. However, beyond a certain point increasing the playback level may not be possible due to loudspeaker overloading or unpleasantly high sound levels~\cite{kleijn_optimizing_2015}. Therefore, many NLE approaches take a power constraint into account, such that equal power is maintained between the unprocessed and processed signal. 

Most work on NLE assumes the signal received from the far-end is noise-free~\cite{kleijn_optimizing_2015}. %That is, the far-end speech signal is not subjected to any noise or other degradations. 
However, in many communication scenarios 
%this assumption is incorrect, such as in mobile phone communication, in which
both the target talker and listener may be in noisy environments. %The subsequent processing to reduce the effects of the far-end interference also influences the speech signal, and the intelligibility at the near-end~\cite{tan_improving_2020}.
Even so, until recently the processing to mitigate the effects of disturbances in the far- and near-end environments have been considered separately. However, recently, in \cite{niermann_joint_2017, khademi_intelligibility_2017, tan_improving_2020} it was shown that optimization of SI under joint consideration of the far- and near-end noise is superior to disjoint processing. 
The work of \cite{niermann_joint_2017} considers jointly controlling a single-channel noise reduction filter along with a post-filter gain for NLE designed to increase SI. % according to the Speech Intelligibility Index (SII)~\cite{american_national_standards_institute_methods_2017}. 
% However, the post-filter is not optimized to maximize SI.
In \cite{tan_improving_2020} a new training strategy is proposed for deep learning based single-channel enhancement given that speech has already been processed at the far-end. For joint multi-microphone FSE and NLE \cite{khademi_intelligibility_2017,khademi_jointly_2016} proposes to optimize the Mutual Information (MI)~\cite{cover_elements_2006} between the clean speech and the signal received by the listener. The results of \cite{khademi_intelligibility_2017} are the first to show both theoretically and experimentally that joint processing, using knowledge of processing and conditions at both ends, is superior to the classic disjoint processing.

MI as a SI optimization objective provides a target that unifies %more classic
heuristic views on SI and 
%newer
mathematically founded SI measures~\cite{khademi_intelligibility_2017,kleijn_simple_2015}. However, solving it in closed form requires simplifying assumptions. The resulting optimization objective is an SNR-type of measure that is approximately equal to the Approximated Speech Intelligibility Index (ASII)~\cite{taal_optimal_2013,khademi_intelligibility_2017}. %[Sec.~VIII.A]
Furthermore, %In addition,
the method of \cite{khademi_intelligibility_2017} depends on the choice of the correlation of the so-called production and interpretation noise with the clean speech.
With the choice made in \cite{khademi_intelligibility_2017}, the objective function of \cite{khademi_intelligibility_2017} reduces fully to the ASII, whereas for more recent choices in \cite{van_kuyk_instrumental_2018} it does not. 

In this paper, we illustrate how, by using a simpler well established signal model and optimizing for the ASII directly, we can derive a closed-form solution to the joint far- and near-end speech enhancement problem that is independent of the marginal distribution of signal coefficients, and without the need for introducing additional parameters in terms of a production and interpretation noise model.
The proposed approach can also be seen as an extension of the ASII optimization problem \cite{taal_optimal_2013} to joint far- and near-end optimization.
Furthermore, we analyze model choices and assumptions 
of \cite{khademi_intelligibility_2017}, and how these relate the approximated MI of \cite{khademi_intelligibility_2017} to the ASII, thus motivating our model choices. Finally, we experimentally compare the performance of \cite{khademi_intelligibility_2017} using the production noise model that was later derived by some of the same authors in \cite{van_kuyk_instrumental_2018, van_kuyk_intelligibility_2016,van_kuyk_information_2017} to our proposed ASII based optimization. We see, that our proposed method achieves similar or slightly better intelligibility in terms of ESTOI~\cite{jensen_algorithm_2016} when the near-end SNR is low and the far-end SNR is intermediate or high.

In summary, the contributions of this paper are: (i) A closed-form solution to ASII based joint far- and near-end speech enhancement, (ii) which is optimal independently of underlying marginal signal distributions, (iii) which does not introduce additional free parameters, e.g., in terms of  production and interpretation noise, and (iv) which performs as good or slightly better than existing schemes.

\section{Existing Work Based on Mutual Information}
\label{sec:existing_work}
MI-based methods for joint far- and near-end SI enhancement \cite{khademi_intelligibility_2017,khademi_jointly_2016} improve SI by maximizing the MI, $I(S;Z)$, between the clean speech, $S$, and the signal received by the listener, $Z$.

\subsection{Existing model assumptions}
In \cite{kleijn_simple_2015}, production and interpretation noise terms are introduced to model natural variations in speakers and listeners, respectively, and adopted into the signal model of \cite{khademi_intelligibility_2017}. The production noise, $Q$, is due to the convolution of the time-domain clean speech and the vocal tract, hence, theoretically, it should be a \emph{multiplicative} noise in the frequency domain.
However, in \cite{khademi_intelligibility_2017} in order to simplify mathematical expressions,
\begin{enumerate}
  \item Multiplicative production noise is modelled as \emph{additive}.
\end{enumerate}
Thus, the single-microphone signal model of \cite{khademi_intelligibility_2017} in the absence of processing is, in the complex short-time DFT (STFT) domain,
\begin{equation}
    Z_{k,i}  = d_{k,i}S_{k,i} + d_{k,i}Q_{k,i} + U_{k,i} + N_{k,i} + W_{k,i},    \label{eq:khadei_model}
\end{equation}
where $k$ is the frequency-bin index and  $i$ the time-frame, $W$ is the interpretation noise, $U$ is the far-end environmental noise, $N$ is the near-end environmental noise, and $d_{k,i}$ are the time-frequency coefficients of the room transfer function from target talker to the microphone. 
The work on MI~\cite{khademi_intelligibility_2017} relies on several common signal model assumptions in the speech processing literature, e.g., that speech and noise STFT coefficients are statistically independent. 
% \begin{enumerate}
%   \item Linear time-invariant processing.
%   \item Stationary and memoryless processes, i.e., they are IID.
%   \item Uncorrelated noise and speech, since they require $\mathop{E}\left[T_k U_k\right] = \mathop{E}\left[T_k \right]\mathop{E}\left[U_k\right]$. However, this requires the assumption of independence, as follows afterwards.
%   \item All processes are independent.
%   \item Independence across frequency.
%   \item The mutual information at the $k$'th frequency is an increasing function of SNR.
  % \item The production and interpretation noise are independent of presentation level and may be represented by a fixed gain at each frequency band, $\rho_{0,k}$.
% \end{enumerate}
However, to derive a speech enhancement procedure based on MI, \cite{khademi_intelligibility_2017} 
introduces additional assumptions and approximations;
\begin{enumerate}
  \setcounter{enumi}{1}
  \item The production noise, $Q$, and interpretation noise, $W$, are independent of the clean speech level and may be represented by a fixed gain (correlation), $\rho_{0,k}$, at each frequency band.
  \item Critical band powers are assumed to be zero-mean independent Gaussian random variables.
\end{enumerate}
In \cite{khademi_intelligibility_2017}, the third assumption is needed since critical band powers are in-fact Chi-squared distributed and the MI between Chi-squared random variables is not expressible in closed form. 
However, we note that critical band powers are positive by definition, hence a zero-mean model is not necessarily appropriate.

\subsection{Approximated Mutual Information vs ASII}\label{sec:asii_mi_comp}
The resulting approximated MI expression in \cite{khademi_intelligibility_2017} is
\begin{equation}
I(S;Z)\approx  -\sum_j\frac{1}{2}\log\left(1-\rho_{0,j}^2\frac{\xi_j}{\xi_j + 1}\right),
\end{equation}
where $\xi_j$ is the SNR in critical band $j$. 
% By \cite[sec. VIII.A]{khademi_intelligibility_2017} and \cite[sec. 2.2]{khademi_jointly_2016} we can write the cost function as
% \begin{equation}
%     I(S; Z) = \sum_j I_j A_j(\xi_j),
% \end{equation}
% where
% \begin{align}
%     A_j(\xi_j) &= \log\left(1-\rho_{0,j}^2\frac{\xi_j}{\xi_j + 1}\right) / \log(1-\rho_{0,j}^2),\\
%     I_j &= -\frac{1}{2}\log(1-\rho_{0,j}^2).
% \end{align}
% This is used in \cite{khademi_jointly_2016,khademi_intelligibility_2017,kleijn_simple_2015} to comment on the similarity between the mutual information measure with production noise, $\rho_0$, and classic intelligibility measures.
% As shown in
By \cite[sec. VIII.A]{khademi_intelligibility_2017} we may take a first order Taylor approximation of the MI in $\rho_{0,j}^2$ around zero, % i.e.,
% \begin{align}
%     A_j(\xi_j) %&= \frac{ \log\left(1-\rho_{0,j}^2\frac{\xi_j}{\xi_j + 1}\right)}{\log(1-\rho_{0,j}^2)} = \frac{\sum_{n=1}^\infty \left(\frac{\rho_{0,j}^2\xi_j}{\xi_j + 1}\right)^n/n}{\sum_{n=1}^\infty \left(\rho_{0,j}^2\right)^n/n}\\
%     &\approx \frac{\frac{\rho_{0,j}^2\xi_j}{\xi_j + 1}}{\rho_{0,j}^2} = \frac{\xi_j}{\xi_j + 1}\label{eq:MI_weight_approx}.
% \end{align}
such that we can approximate the cost function as,
\begin{equation}
    I(S;Z)\approx \sum_j I_j  \frac{\xi_j}{\xi_j + 1},\label{eq:MI_weight_approx}
\end{equation}
for $\rho_{0,j}^2 \to 0^+$, where $I_j \triangleq -\frac{1}{2}\log(1-\rho_{0,j}^2)$.
In the absence of a production noise model at the time, the values for $\rho_{0,j}$ where derived in \cite{khademi_intelligibility_2017} based on the band importance functions, $\gamma_j$, of the SII~\cite{american_national_standards_institute_methods_2017} such that 
$\rho_{0,j}^2 = {1-2^{-2\gamma_j}}$. 
Inserting this, the cost function is
\begin{equation}
    I(S;Z)\approx \sum_j -\frac{1}{2}\log(1-(1-2^{-2\gamma_j}))  \frac{\xi_j}{\xi_j + 1}.\label{eq:MI_apx_full}
\end{equation}
We recognize,  \eqref{eq:MI_apx_full} resembles the ASII introduced in \cite{taal_optimal_2013} as a cost function for SI enhancement. The ASII is defined as,
\begin{equation}
    ASII \triangleq \sum_{j}\gamma_j f(\xi_j),~\label{eq:asii_def} \quad
    % \\
    f(\xi_j)\triangleq \frac{\xi_j}{\xi_j + 1},%\label{eq:sii_apx_fxi}
\end{equation}
where the weights $\gamma_j$ are the critical-band importance functions as defined in \cite{american_national_standards_institute_methods_2017}, 
and $f(\xi_j)$ is the audibility function per critical band.
We notice from \cite[Table~1]{american_national_standards_institute_methods_2017} that band importance functions, $\gamma_j$, are in the interval of $[0.01, 0.06]$, resulting in $\rho_{0,j}^2 \in [0.0138, 0.0798]$. As shown in \cite[Fig. 1]{khademi_jointly_2016} the approximation \eqref{eq:MI_weight_approx} holds for ${\rho_{0,j}^2 \leq 0.4}$. Hence, we can conclude the choice of $\rho_{0,j}^2= 1-2^{-2\gamma_j}$ to be sufficiently close to zero for equality to hold in \eqref{eq:MI_apx_full}.
Thus, the MI problem in \cite{khademi_intelligibility_2017} is equal to the ASII problem, when the parameter modelling production- and interpretation noise, $\rho_{0,j}$, is chosen according to the band importance functions of the SII. 
%  Therefore, we consider optimization of ASII directly using a simpler signal model without natural speech variations and the assumptions of previous work.

\section{Signal Model}\label{sec:signal_model}
\begin{figure}[tb]
    \centering
    % Definition of blocks:
\tikzset{%
  block/.style    = {draw, thick, rectangle, minimum height = 3em,
    minimum width = 3em, align=center},
  sum/.style      = {draw, circle, node distance = 2cm}, % Adder
  input/.style    = {coordinate}, % Input
  output/.style   = {coordinate}, % Output
  cross/.style={path picture={ 
	\draw[black]
  (path picture bounding box.south east) -- (path picture bounding box.north west) (path picture bounding box.south west) -- (path picture bounding box.north east);}},
  circle_gain/.style = {draw, cross, circle, minimum height=2.3em}
}
% Defining string as labels of certain blocks.
\newcommand{\suma}{\Large$+$}
\newcommand{\inte}{$\displaystyle \int$}
\newcommand{\derv}{\huge$\frac{d}{dt}$}

\begin{tikzpicture}[auto, thick, node distance=2.5cm, >=triangle 45]
% \draw[help lines] (-3,-6) grid (11,2);
% \draw[step=1mm, gray, thin] (0,0) grid (5,5); 
\draw
	% Drawing the blocks of first filter :
	node at (0,0)[left=-3mm] (source) { $\bm{S}_k$}
	% node [sum, right of=source] (prod_noise_sum) {\suma}
	% node [above of=prod_noise_sum] (prod_noise) {$Q$}
	
	node [circle_gain, right of=source, node distance=1.5cm] (IR_mult) {}
	node [above of=IR_mult, node distance=1.5cm] (IR) {$\bm{d}_k$}
	
	node [sum, right of=IR_mult] (FE_noise_sum) {\suma}
	node [above of=FE_noise_sum, node distance=1.5cm] (FE_noise) {$\bm{U}$}
	node [block, right of=FE_noise_sum, node distance=2.3cm] (mvdr) { $\bm{w}_k$}
	node[circle_gain, right of=mvdr, node distance=2.3cm] (freq_gain) {}
	node[above of=freq_gain, node distance=1.5cm] (alpha) {$\sqrt{\alpha_k}$}
	% node [sum, right of=freq_gain, node distance=3cm] (code_noise_sum) {\suma}
	% node [above of=code_noise_sum] (code_noise) {$Q_k$}
	% node [block, right of=code_noise_sum, node distance=2.7cm] (post_proc) {Post-\\Processing}
	node [sum, right of=freq_gain, node distance=2.3cm] (NE_noise_sum) {\suma}
	node [above of=NE_noise_sum, node distance=1.5cm] (NE_noise) {$N_k$}
	% node [sum, right of=NE_noise_sum] (intp_noise_sum) {\suma}
	% node [above of=intp_noise_sum] (intp_noise) {$W$}
	node [right of=NE_noise_sum, node distance=1.5cm] (output) {$Z_k$}
	;
    % Joining blocks. 
    % Commands \draw with options like [->] must be written individually
	\draw[->](source) -- node {}(IR_mult);
	% \draw[->](prod_noise) -- (prod_noise_sum);
 	% \draw[->](prod_noise_sum) --node[above]{$T$} (IR_mult);
 	\draw[->](IR) -- (IR_mult);
 	\draw[->](IR_mult) -- (FE_noise_sum);
 	\draw[->](FE_noise) -- (FE_noise_sum);
	\draw[->](FE_noise_sum) -- node[above, pos=0.4]{$X_k$} (mvdr);
    \draw[->](mvdr) -- (freq_gain);
    \draw[->](alpha) -- (freq_gain);
	\draw[->](freq_gain) -- node[above]{$Y_k$} (NE_noise_sum);
	% \draw[->](code_noise) -- (code_noise_sum);
    % \draw[->](code_noise_sum) --node[above]{$\widehat{Y}$}  (post_proc);
	% \draw[->](code_noise_sum) --node[above]{\Large$\widetilde Y_k$} (NE_noise_sum);
	\draw[->](NE_noise) -- (NE_noise_sum);
	% \draw[->](NE_noise_sum) -- node[above]{$Y$}(intp_noise_sum);
	% \draw[->](intp_noise) -- (intp_noise_sum);
	\draw[->](NE_noise_sum) -- node{}(output);

\node[color=gray,thick, dashed, draw, rectangle, xshift=.5cm, yshift=.6cm,minimum width=3.9cm, minimum height=2.5cm, label={Processing}] at (mvdr.east) {};
\end{tikzpicture}
    % !!! Blurry in PDF preview only not in Adobe !!!
    \caption{\normalsize Our signal model of optimal joint SI enhancement.}
    \label{fig:optimal_block_diagram}
\end{figure}
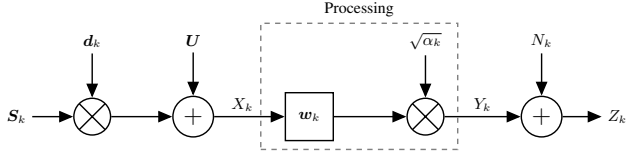

In this section we introduce the proposed signal model, cf. Fig.~\ref{fig:optimal_block_diagram}. 
The single-microphone signal model follows,
% \begin{subequations}
%     \begin{align}
%         X_{k,i} &= d_{k,i}S_{k,i} + U_{k,i},\\
%         Y_{k,i} &= vX_{k,i},\\
%         Z_{k,i} &= Y_{k,i} + N_{k,i},
%     \end{align}
% \end{subequations}
\begin{align}
    X_{k,i}= d_{k,i}S_{k,i} + U_{k,i},~Y_{k,i}= vX_{k,i}, ~Z_{k,i}= Y_{k,i} + N_{k,i},
\end{align}
where $X_{k,i}$ is the recorded signal in STFT domain, i.e., the clean speech, $S_{k,i}$, recorded by the microphone contaminated by the far-end noise, $U_{k,i}$. 
To increase SI of the received message, the noisy microphone signal, $X_{k,i}$, is linearly processed prior to playout, producing the modified signal $Y_{k,i}$. %Finally, we assume the linear processor, $v$, is derived while aware of relevant side information, $\mathcal{S}_U$, $\mathcal{S}_N$, about the far-end noise and near-end noise.
The signal received by the listener, $Z_{k,i}$, is finally contaminated by the noise in the near-end environment, $N_{k,i}$. 
The speech and noise processes, $S$, $U$, and $N$, are assumed to be stationary sequences of complex random vectors consisting of the STFT coefficients. Both the far-end noise, $U$, and the near-end noise, $N$ are assumed to be independent of each other and of the target speech, $S$.
 These assumptions are similar to \cite{khademi_intelligibility_2017}. However, compared to \cite{khademi_intelligibility_2017} we do not need to model a multiplicative production noise as additive, or to introduce an additional interpretation noise. Further, we do not need assumptions on the particular marginal distributions of the signals.
% \begin{figure}[tb]
%     \centering
%     \includegraphics[width=\columnwidth]{system_model_pre_process_side-info.pdf}
%     \caption{\normalsize Our system model.}
%     \label{fig:our_sys}
% \end{figure}

\subsection{Multi-Microphone Signal Model}
Let us denote the acoustic transfer function from source to microphone $m$ by $d_{k,i,m}$ with vector notation, $\bm{d}_{k,i} = \left[d_{k,i,1}, \ldots, d_{k,i,m}\right]^T$, and letting vector $\bm{U}_{k,i}$ denote far-end noise recorded by the microphones. Then the noisy microphone signals are given by
\begin{equation}
    \bm{X}_{k,i} = \bm d_{k,i} S_{k,i} + \bm{U}_{k,i}.
\end{equation} 
Denoting the linear multi-microphone processor by $\bm{v}_{k,i}$, the processed microphone signal is, 
\begin{align}
    Y_{k,i} &= \bm{v}_{k,i}^H \bm d_{k,i} S_{k,i} + \bm{v}_{k,i}^H \bm{U}_{k,i}
    % &= \widetilde{S}_{k,i}+ \widetilde{U}_{k,i},
\end{align}
where super-script $H$ denotes conjugate transpose.

\section{Optimal ASII Linear Processor}\label{sec:asii_opt}
In this section we derive the optimal linear processor based on the ASII defined in~\eqref{eq:asii_def}. The derivation steps are similar to \cite{khademi_intelligibility_2017} and \cite{taal_optimal_2013}. However, contrary to \cite{khademi_intelligibility_2017} we consider the ASII instead of MI. Further, we expand on \cite{taal_optimal_2013} by including joint (multi-microphone) processing with far-end noise.

The energy of the clean speech signal within one critical band, $j$, and time-frame, $i$, is defined as
\begin{equation}
    \mathcal{S}_{j,i}^2 \triangleq \sum_k \lvert S_{k,i} \rvert^2 \lvert H_j(k)\rvert^2,
\end{equation}
where $H_j(k)$ is the STFT coefficients of the $j$'th critical band filter. Similarly, we define the critical band energy of the near-end noise, $\mathcal{N}_{j,i}^2$, and the processed far-end speech, $\mathcal{\widetilde{S}}_{j,i}^2$, and noise, $\mathcal{\widetilde{U}}_{j,i}^2$.
%  within one critical band as; 
% $
% % {\mathcal{Q}_{j,i}^2 \triangleq \sum_k \lvert Q_{k,i} \rvert^2 \lvert H_j(k)\rvert^2},~
% {\mathcal{N}_{j,i}^2 \triangleq \sum_k \lvert N_{k,i} \rvert^2 \lvert H_j(k)\rvert^2},$ 
% $
% {\mathcal{\widetilde{U}}_{j,i}^2 \triangleq \sum_k \lvert \widetilde{U}_{k,i} \rvert^2 \lvert H_j(k)\rvert^2},$ and
% $
% {\mathcal{\widetilde{S}}_{j,i}^2 \triangleq \sum_k \lvert \widetilde{S}_{k,i} \rvert^2 \lvert H_j(k)\rvert^2},
% $
% % \begin{align}
% %     % \mathcal{Y}_{m,i}^2 &\triangleq \sum_k \lvert Y_{k,i} \rvert^2 \lvert H_j(k)\rvert^2,\\
% %     \mathcal{Q}_{j,i}^2 &\triangleq \sum_k \lvert Q_{k,i} \rvert^2 \lvert H_j(k)\rvert^2,\\
% %     \mathcal{N}_{j,i}^2 &\triangleq \sum_k \lvert N_{k,i} \rvert^2 \lvert H_j(k)\rvert^2,\\
% %     \mathcal{\widetilde{U}}_{j,i}^2 &\triangleq \sum_k \lvert \widetilde{U}_{k,i} \rvert^2 \lvert H_j(k)\rvert^2,\\
% %     \mathcal{\widetilde{S}}_{j,i}^2 &\triangleq \sum_k \lvert \widetilde{S}_{k,i} \rvert^2 \lvert H_j(k)\rvert^2,
% % \end{align}
% where ${\widetilde{U}_{k,i} = \bm{v}_{k,i}^H \bm{U}_{k,i}}$ is the processed far-end noise, and ${\widetilde{S}_{k,i} = \bm{v}_{k,i}^H \bm{d}_{k,i} {S}_{k,i}}$ is the processed speech signal.
Since we assume stationarity of the speech and noise, we can disregard the time-index, $i$, and let the average energy per DFT bin and critical band be based on a long-term average over several short-time frames,
\begin{equation}
    \sigma_{S_k}^2 \triangleq \frac{1}{I}\sum_i  \lvert S_{k,i} \rvert^2, \quad \sigma_{\mathcal{S}_j}^2\triangleq \sum_k \lvert H_j(k)\rvert^2\sigma_{S_k}^2,
\end{equation}
where $I$ is the total number of frames. 
% Then the average energy within one critical band is defined as
% \begin{equation}
%     \sigma_{\mathcal{S}_j}^2\triangleq \sum_k \lvert H_j(k)\rvert^2\sigma_{S_k}^2 .
% \end{equation}
Similar definitions hold for the noise terms $U$ and $N$. The critical band filters, $H_{j}(k)$, are normalized such that the total energy is preserved in critical bands, i.e.,
\begin{equation}
   \sum_j \sigma_{\mathcal{S}_j}^2 = \sum_k \sigma_{{S}_k}^2. \label{eq:cb_dft_ener_eq}
\end{equation}
The critical band SNR at the near-end listener is then,% given as
\begin{align}
    \xi_j &= \frac{\sigma_{\widetilde{\mathcal{S}}_j}^2}{  \sigma_{\widetilde{\mathcal{U}}_j}^2 +  \sigma_{\mathcal{N}_j}^2}.
\end{align}
Inserting this into \eqref{eq:asii_def}, we have that
\begin{align}
    f(\xi_j) %&=
    %\tfrac{\sum_k\lvert H_j(k)\rvert^2 \sigma_{\widetilde{S}_k}^2}{  \sum_k\lvert H_j(k)\rvert^2 \left(\sigma_{\widetilde{S}_k}^2+  \sigma_{\widetilde{{U}}_{k}}^2 +  \sigma_{{N}_{k}}^2\right)}\\
    &=\tfrac{\sum_k\lvert H_j(k)\rvert^2\bm{v}_{k}^H \bm{d}_{k} \bm{d}_{k}^H \bm{v}_{k} \sigma_{S_{k}}^2}{ \sum_k\lvert H_j(k)\rvert^2\left( \bm{v}_{k}^H \bm{d}_{k} \bm{d}_{k}^H \bm{v}_{k} \sigma_{S_{k}}^2+  \bm{v}_{k}^H \Sigma_{U_{k}} \bm{v}_{k} + \sigma_{{N}_{k}}^2\right)}\\
    &\triangleq  f_j\left(\left\{\bm{v}_k, \Theta_k \right\} \right),
\end{align}
where $\Theta_k = (\sigma_{S_{k}}^2, \Sigma_{U_{k}}, \sigma_{{N}_{k}}^2)$. 
In order to limit loudspeaker overload or unpleasant playback levels we invoke the following equal power constraint,
 \begin{equation}
    \sum_k \bm{v}_{k,i}^H \bm{d}_{k,i} \bm{d}_{k,i}^H \bm{v}_{k,i} \sigma_{S_{k,i}}^2 = \sum_{k} \sigma_{S_{k,i}}^2.
 \end{equation}
 That is, for each time frame $i$ the total power of the clean speech is unaltered by processing. 
The joint far and near-end SI enhancement problem with equal power constraint is then,
\begin{equation}
    \begin{array}{ll}
        \displaystyle \mathop{\sup}_{\{\bm{v}_k\} \in \mathbb{C}^M} & 
        \sum_j\gamma_j f_j\left(\left\{\bm{v}_k, \Theta_k\right\} \right)
        \\
        \mbox{subject to} & \sum_k \bm{v}_k^H \bm{d}_k \bm{d}_k^H \bm{v}_k \sigma_{S_k}^2 = \sum_k \sigma_{S_k}^2.
    \end{array}
\end{equation}
We now introduce the real and positive variable, $\alpha_k$, to perform a variable transformation $\bm{v}_k = \alpha_k^{1/2}\bm{w}_k$ with the additional constraint $\bm{v}_k^H \bm{d}_k \bm{d}_k^H \bm{v}_k = \alpha_k$. This leads to the equivalent problem,
\begin{equation}
    \begin{array}{ll}
        \displaystyle \mathop{\sup}_{\{\bm{v}_k\} \in \mathbb{C}^M, \{\alpha_k\}\in \mathbb{R}_+} & 
        \sum_j\gamma_j f_j\left(\left\{\alpha_k, \bm{v}_k, \Theta_k\right\} \right)
        \\
        \mbox{subject to} & \mathcal{C}_1~:~\sum_k \alpha_k \sigma_{S_k}^2 = \sum_k \sigma_{S_k}^2,\\
        &  \mathcal{C}_2~:~\bm{v}_k^H \bm{d}_k \bm{d}_k^H \bm{v}_k = \alpha_k,~\forall k .
    \end{array}
\end{equation}
The objective function can be rewritten in terms of $\alpha_k$ and $\bm{w}_k$, i.e.,
\begin{align*}
    f_j\left(\left\{\alpha_k, \bm{w}_k, , \Theta_k\right\} \right)=
    \tfrac{\sum_k\lvert H_j(k)\rvert^2 \alpha_k \sigma_{S_{k}}^2}{ \sum_k\lvert H_j(k)\rvert^2 \left( \alpha_k \sigma_{S_{k}}^2+   \alpha_k \bm{w}_{k}^H \Sigma_{U_{k}} \bm{w}_{k}  +   \sigma_{{N}_{k}}^2\right)}.
\end{align*}
We notice that $\bm{v}_k^H \bm{d}_k \bm{d}_k^H \bm{v}_k = \alpha_k \Leftrightarrow \bm{d}_k^H \bm{w}_k = 1.$ Hence, writing the optimization problem in terms of $\bm{w}_k$ and $\alpha_k$ we have,
\begin{equation}
    \begin{array}{ll}
        \displaystyle \mathop{\sup}_{\{\bm{w}_k\} \in \mathbb{C}^M, \{\alpha_k\}\in \mathbb{R}_+} &   \sum_j\gamma_j f_j\left(\left\{\alpha_k, \bm{w}_k,  \Theta_k\right\} \right)\\
        \mbox{subject to} & \mathcal{C}_1~:~\sum_k \alpha_k \sigma_{S_k}^2 = \sum_k \sigma_{S_k}^2\\
        &  \mathcal{C}_2~:~ \bm{d}_k^H \bm{w}_k = 1. \label{eq:ASII_opt_var_trans}
    \end{array}
\end{equation}
% Using the relation $\sup_{x,y}f(x,y)=\sup_x \sup_y f(x,y)$~\cite[p.~133]{boyd_convex_2004}, 
We can separate \eqref{eq:ASII_opt_var_trans} across the two variables~\cite[p.~133]{boyd_convex_2004}, i.e.,% and their independent constraints,
\begin{equation*}
        \displaystyle 
        \mathop{\sup}_{\{\alpha_k\}\in \mathbb{R}_+, \mathcal{C}_1}
        \mathop{\sup}_{\{\bm{w}_k\} \in \mathbb{C}^M, \mathcal{C}_2} \sum_j\gamma_jf_j\left(\left\{\alpha_k, \bm{w}_k, \Theta_k \right\} \right).
\end{equation*}
The inner optimization problem across $\{\bm{w}_k\}$ corresponds to the standard Minimum Variance Distortionless Response (MVDR) beamforming problem with the solution~\cite{gannot_consolidated_2017},
\begin{equation}
    \bm{w}_{k}^* = \frac{\Sigma_{U_{k,i}}^{-1}\bm{d}_{k,i}}{\bm{d}_{k,i}^H\Sigma_{U_{k,i}}^{-1}\bm{d}_{k,i}},\quad \forall k.
\end{equation}
% This spatial processing using the MVDR beamformer is similar to the spatial processing of \cite{khademi_intelligibility_2017}.
Inserting the MVDR solution into \eqref{eq:ASII_opt_var_trans}, the remaining problem is
\begin{equation}
    \begin{array}{ll}
        \displaystyle \mathop{\sup}_{\{\alpha_k\}\in \mathbb{R}_+} &   \sum_j\gamma_j \left(\frac{\sum_k\lvert H_j(k)\rvert^2 \alpha_k \sigma_{S_{k}}^2}{  \sum_k\lvert H_j(k)\rvert^2 \left(\alpha_k \sigma_{S_{k}}^2+ \alpha_k \sigma_{B_k}^2 + \sigma_{{N}_{k}}^2\right)} \right)\\
        \mbox{subject to} & \mathcal{C}_1~:~\sum_k \alpha_k \sigma_{S_k}^2 = \sum_k \sigma_{S_k}^2, \label{eq:ASII_opt_remain}
    \end{array}
\end{equation}
where $\sigma_{B_k}^2 \triangleq \bm{w}_{k}^{*^H} \Sigma_{U_{k}} \bm{w}_{k}^*$ is the residual far-end noise after processing by the MVDR beamformer. 

\subsection{Critical-band near-end optimization}\label{sec:asii_crit_band}
We derive, similarly to existing work on SI enhancement \cite{khademi_intelligibility_2017,niermann_joint_2017,taal_optimal_2013,kleijn_optimizing_2015}, the optimal near-end processor based on the assumption that all frequency gains within a critical band $j$ are the same, i.e.,
\begin{equation}
    \alpha_k = \alpha_{k'}, \forall k, k' \in \mathcal{K}_j, 
\end{equation}
where $\mathcal{K}_j$ is the set of frequency bins belonging to the $j$'th critical band. The gains are then later on converted back to DFT domain.

Starting from the optimization problem \eqref{eq:ASII_opt_remain} we have,
\begin{equation}
    \begin{array}{ll}
        \displaystyle \mathop{\sup}_{\{\alpha_j\}\in \mathbb{R}_+} &   \sum_j\gamma_j \frac{\alpha_j \sigma_{\mathcal{S}_j}^2}{  \alpha_j \sigma_{\mathcal{S}_j}^2+   \alpha_j \sigma_{\mathcal{B}_j}^2 + \sigma_{\mathcal{N}_j}^2} \\
        \mbox{subject to} & \mathcal{C}_1~:~\sum_j \alpha_j \sigma_{\mathcal{S}_j}^2 = \sum_j \sigma_{\mathcal{S}_j}^2. \label{eq:ASII_crit_opt_remain}
    \end{array}
\end{equation}
We notice each term in the sum is concave in $\alpha_j$ for $\alpha_j\geq0$.
Therefore, the weighted sum of these concave functions is also concave. 
We describe the problem by the Lagrangian cost-function~\cite{boyd_convex_2004},
\begin{align}
    \mathcal{L} =  &\sum_j\tfrac{\gamma_j \alpha_j \sigma_{\mathcal{S}_j}^2}{  \alpha_j \sigma_{\mathcal{S}_j}^2+   \alpha_j \sigma_{\mathcal{B}_j}^2 + \sigma_{\mathcal{N}_j}^2}  \nonumber
    - \nu \left(\sum_j \alpha_j \sigma_{\mathcal{S}_j}^2 - r\right)
    + \sum_j \lambda_j \alpha_j,
\end{align}
where $r = \sum_j  \sigma_{\mathcal{S}_j}^2$, and $\nu$ and $\lambda_j$ are the Lagrangian multipliers for the energy constraint and inequality constraint in  \eqref{eq:ASII_crit_opt_remain}. 
The KKT conditions~\cite{boyd_convex_2004} for the optimization problem are,
\begin{subequations}
    \begin{align}
        r&=\sum_j \alpha_j \sigma_{\mathcal{S}_j}^2 \label{eq:asii_crit_kkt_energy},~
        0\leq \alpha_j ,~ %\forall j,
        0\leq \lambda_j ,~% \forall j
        0=\lambda_j \alpha_j , \forall j,\\% \label{eq:asii_crit_kkt_comp_slack}
        0&=\gamma_j\frac{\sigma_{\mathcal{S}_j}^2 \sigma_{\mathcal{N}_j}^2}{\left(  \alpha_j \left(\sigma_{\mathcal{S}_j}^2+  \sigma_{\mathcal{B}_j}^2\right) + \sigma_{\mathcal{N}_j}^2 \right)^2}
         - \nu  \sigma_{\mathcal{S}_j}^2 
         + \lambda_j, \forall j \label{eq:asii_crit_kkt_grad}
    \end{align}
\end{subequations}
Isolating $\lambda_j$ in \eqref{eq:asii_crit_kkt_grad}, then using the complimentary slackness condition %\eqref{eq:asii_crit_kkt_comp_slack}  
to set $\lambda_j = 0$, we solve for the non-zero $\alpha_j$,
\begin{equation}
    \alpha_j = \max \left\{\frac{\sqrt{\sigma_{\mathcal{N}_j}^2\gamma_j}}{\sqrt{\nu} \left(\sigma_{\mathcal{S}_j}^2+  \sigma_{\mathcal{B}_j}^2\right)} - \frac{  \sigma_{\mathcal{N}_j}^2}{\sigma_{\mathcal{S}_j}^2+  \sigma_{\mathcal{B}_j}^2} , 0 \right\},~\forall j, \label{eq:asii_crit_alpha}
\end{equation}
where $\nu$ is chosen such that the energy constraint in \eqref{eq:asii_crit_kkt_energy} is satisfied,
\begin{align}
     \frac{1}{\sqrt{\nu}}
     &= 
    %  \frac{
    %     r + \sum\limits_{j \in \mathcal{J}} \frac{\sigma_{\mathcal{S}_j}^2 (\sigma_{\mathcal{N}_j}^2)}{\sigma_{\mathcal{S}_j}^2+  \sigma_{\mathcal{B}_j}^2}
    %      }
    %      {
    %     \sum\limits_{j \in \mathcal{J}}  \left(\frac{ \sigma_{\mathcal{S}_j}^2\sqrt{  \sigma_{\mathcal{N}_j}^2}\sqrt{\gamma_j }}{ \left(\sigma_{\mathcal{S}_j}^2+  \sigma_{\mathcal{B}_j}^2\right)} \right)
    %          },\label{eq:asii_crit_nu}
    \left(
    r + \sum\limits_{j \in \mathcal{J}} \tfrac{\sigma_{\mathcal{S}_j}^2 \sigma_{\mathcal{N}_j}^2}{\sigma_{\mathcal{S}_j}^2+  \sigma_{\mathcal{B}_j}^2}
    \right)/
    \left(
    \sum\limits_{j \in \mathcal{J}}  \left(\tfrac{ \sigma_{\mathcal{S}_j}^2\sqrt{  \sigma_{\mathcal{N}_j}^2\gamma_j} }{ \left(\sigma_{\mathcal{S}_j}^2+  \sigma_{\mathcal{B}_j}^2\right)} \right)
    \right),\label{eq:asii_crit_nu}
\end{align}
here $\mathcal{J} = \left\{j \in \mathbb{N} : \alpha_j > 0\right\}$ denotes the set of frequency bins for which the optimal $\alpha_j$ are positive. We notice, that the set of indices $\mathcal{J}$ depends on the $\alpha_j$\cite{taal_optimal_2013}. Therefore, $\nu$ also depends on $\alpha_j$. Hence, there is a recursive relationship between \eqref{eq:asii_crit_alpha} and \eqref{eq:asii_crit_nu}, which may be resolved by using, e.g., a bi-section method or evaluating \eqref{eq:asii_crit_alpha} for a range of $\nu$ values, such that the energy constraint is satisfied~\cite{taal_optimal_2013}.
Finally, to get the optimal frequency dependent gains, $\alpha_k^*$, we weight the optimal, $\alpha_j^*$, using the critical band filters, that is
\begin{equation}
    \alpha_k^* = \sum_j \lvert H_j(k)\rvert^2 \alpha_j^*,
\end{equation}
where the energy constraint is satisfied in both frequency and critical-band domain as per the normalization of the critical band filters in \eqref{eq:cb_dft_ener_eq}, i.e., $\sum_j  \alpha_j^* \sigma_{\mathcal{S}_j}^2 = \sum_j \sigma_{\mathcal{S}_j}^2= \sum_k \sigma_{{S}_k}^2 =  \sum_k \alpha_k^* \sigma_{{S}_k}^2.$
% \begin{equation}
%   \sum_j  \alpha_j^* \sigma_{\mathcal{S}_j}^2 = \sum_j \sigma_{\mathcal{S}_j}^2= \sum_k \sigma_{{S}_k}^2 =  \sum_k \alpha_k^* \sigma_{{S}_k}^2.
% \end{equation}

The proposed processor is summarized in Fig.~\ref{fig:optimal_block_diagram}. As in \cite{khademi_intelligibility_2017}, the procedure consists of an MVDR beamformer, $\bm{w}_k$, followed by a frequency dependent gain, $\alpha_k$. %This is similar to the results of \cite{khademi_intelligibility_2017}, where it is also mentioned how the MVDR is the optimal first operation for any intelligibility measure that is an increasing function of SNR.
 In contrast to the results of \cite{khademi_intelligibility_2017}, the frequency bin gains, $\alpha_k$, of our procedure are optimized specifically according to the ASII without approximations and assumptions on signal distributions and without introduction of additional free parameters to model natural speech variations.

\section{Experimental Evaluation}\label{sec:exp_eval}
We have seen (Section~\ref{sec:existing_work}) that by using the production noise model choice in  \cite{khademi_intelligibility_2017} the MI cost function reduces to the ASII. However, this choice of production noise is due to a lack of a more accurate model at the time. Recently, some of the authors have provided an estimated model for the production noise in \cite{van_kuyk_instrumental_2018, van_kuyk_intelligibility_2016,van_kuyk_information_2017}, where $\rho_{0,j}=0.75~\forall j$. 
We compare performance of our proposed method with \cite{khademi_intelligibility_2017} using the newer production noise model of \cite{van_kuyk_instrumental_2018}. We consider a Python simulation of a setup similar to that of \cite{khademi_intelligibility_2017, khademi_joint_2017_code}. 

\subsection{Experimental Setup}
We consider a room with dimensions $3 \times 4 \times 3$\,\si{m^3}, a single target speaker located at $[1.50, 3.00, 1]$\,\si{m}, and an array with two microphones spaced $\SI{2}{cm}$ apart at $[1.50, 2.00, 1]$\,\si{m} and $[1.50, 2.02, 1]$\,\si{m}. At the far-end there are three speech shaped noise sources  located at $[0.50, 1.00, 1]$\,\si{m}, $[0.75, 3.00, 1]$\,\si{m} and $[3.00, 1.60, 1]$\,\si{m}, respectively. The near-end noise is pink. In addition to the far-end noise each microphone is subject to a $\SI{60}{dB}$ SNR  white noise. The source signal is speech signals from five female and five male speakers from the TIMIT-database~\cite{timit_1993} sampled at $\SI{16}{kHz}$.
Signals were processed block-wise based on a DFT with $\SI{32}{ms}$ Hann windows with $50\%$ overlap. Since we assume stationarity, the MVDR beamformer, $\bm{w}_k$ and post-filter gains, $\alpha_k$, are derived based on the long-term average spectrums, leading to time-invariant processing.
The long-term power of the clean speech, far-end noise, and the near-end noise are assumed to be known. Hence, we do not estimate any of these spectrums. Furthermore, the room transfer functions are assumed to be known, and generated without reverberation using \cite{allen_image_1979}.% implemented in \cite{yoshioka_rir-generator_2019}.

\subsection{Results}
\begin{figure}[tb]
    \centering
    \includegraphics[width=\columnwidth]{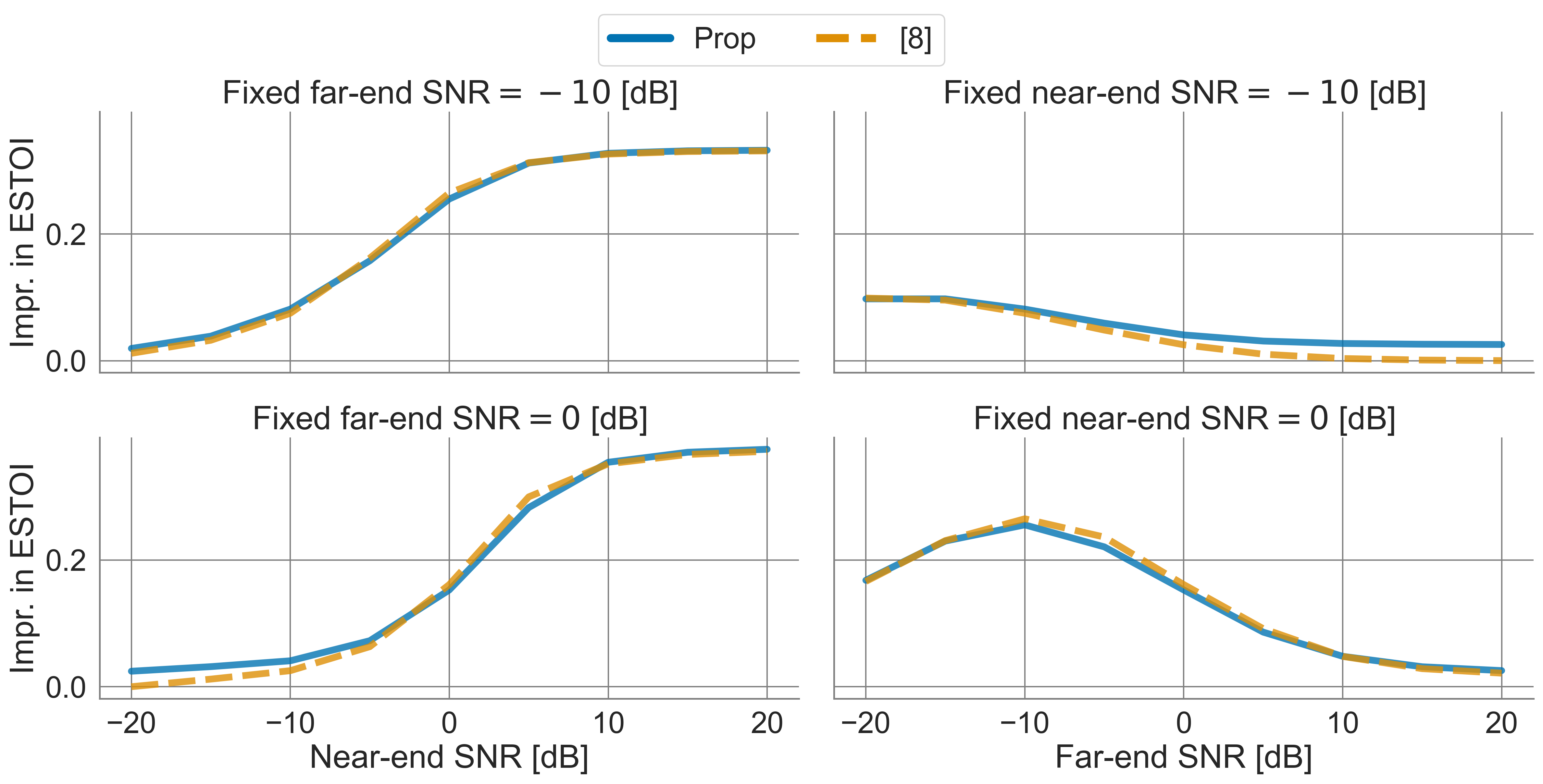}
    \caption{\normalsize ESTOI performance of the proposed method and~\cite{khademi_intelligibility_2017}, for varying near-end SNR and a fixed far-end SNR (left column), and varying far-end SNR and a fixed near-end SNR (right column).    
    }
    \label{fig:asii_estoi_perform}
\end{figure}

Fig.~\ref{fig:asii_estoi_perform} shows improvements over 'unprocessed' in ESTOI~\cite{jensen_algorithm_2016}, of the proposed method along with the method of \cite{khademi_intelligibility_2017}. 
The results show that generally the two methods achieve a similar performance, which is expected due to the similarity shown in Section~\ref{sec:existing_work}. However, the proposed method is slightly better when the near-end SNR is low and the far-end SNR is intermediate or high. Thus, the production noise model choice of \cite{van_kuyk_instrumental_2018} leads to a slightly worse speech enhancement than using the model based on the SII weights~\cite{american_national_standards_institute_methods_2017}. ASII plots show identical performance between the two methods and are thus not reported.

\section{Conclusion}
We have derived a closed-form optimal linear processor for joint far- and near-end speech intelligibility enhancement based on ASII. The optimal processor consists of an MVDR beamformer followed by a frequency dependent post gain. The derived processor is based on a simple model without relying on assumptions and approximations of the underlying marginal signal distributions. Furthermore, we do not need to model or optimize for natural variations in speech. Finally, as a consequence, the proposed processor has comparable or slightly better ESTOI performance than existing work.  %The paper also highlights how existing work is similar to ours because of a lack of a more thorough model for production and interpretation noise, thus leading to a mutual information expressions that is equivalent to the ASII.

% -------------------------------------------------------------------------

\bibliographystyle{IEEEbib}
% \bibliography{../../Worksheets/bib/phd_bib.bib}
\bibliography{ICASSP2022_bib.bib}

\end{document}